%% file: 0-HexGlyph.tex
\documentclass[preprint]{vgtc}               

\ifpdf
  \pdfoutput=1\relax                   
  \usepackage{graphicx}                
  \DeclareGraphicsExtensions{.pdf,.png,.jpg,.jpeg} 
\else
  \usepackage{graphicx}                
  \DeclareGraphicsExtensions{.eps}     
\fi%

\graphicspath{{figures/}{pictures/}{images/}{./}} 

\usepackage{microtype}                 
\PassOptionsToPackage{warn}{textcomp}  
\usepackage{textcomp}                  
\usepackage{mathptmx}                  
\usepackage{times}                     
\usepackage{cite}                      
\usepackage{tabu}                      
\usepackage{booktabs}                  

\usepackage{xcolor}
\usepackage{color, colortbl}

\definecolor{myblue}{rgb}{0,0,0}
\newcommand{\revise}[1]{{\textcolor{myblue}{#1}}}

\usepackage{wrapfig}
\usepackage{subfig}

\onlineid{1053}

\vgtccategory{Tools}

\vgtcinsertpkg

\title{A Visualization System for Hexahedral Mesh Quality Study}

\author{Lei Si\thanks{e-mail: lsi@uh.edu}\\ %
        \scriptsize University of Houston%
\and Guoning Chen\thanks{e-mail: gchen16@uh.edu}\\ %
\scriptsize University of Houston %
}

\input{contents/1-Abstract}

\CCScatlist{
  \CCScatTwelve{Human-centered computing}{Visu\-al\-iza\-tion}{Visualization systems and tools}
}


\begin{document}


\setlength{\baselineskip}{0.99 \baselineskip}

\setlength{\abovedisplayskip}{0pt}
\setlength{\belowdisplayskip}{0pt}
\setlength{\abovedisplayshortskip}{3pt}
\setlength{\belowdisplayshortskip}{-2pt}
\setlength{\belowcaptionskip}{-2pt}
\setlength{\abovecaptionskip}{0pt}
\setlength{\textfloatsep}{5pt}
\setlength{\floatsep}{3pt}
\setlength{\intextsep}{2pt}

\maketitle

\input{contents/2-Introduction}

\input{contents/3-RelatedWord}
\input{contents/4-OurMethod}
\input{contents/6-UserCaseAndEvaluation}
\input{contents/7-Conclusion}
\acknowledgments{
We wish to thank the anonymous reviews for their constructive feedback to help improve this work.}
\bibliographystyle{abbrv-doi}
\bibliography{contents/HQView-bib}
\end{document}

%% file: contents/1-Abstract.tex
\abstract{
In this paper, we introduce a new 3D hex mesh visual analysis system that emphasizes poor-quality areas with an aggregated glyph, highlights overlapping elements, and provides detailed boundary error inspection in three forms. By supporting multi-level analysis through multiple views, our system effectively evaluates various mesh models and compares the performance of mesh generation and optimization algorithms for hexahedral meshes.
}

%% file: contents/2-Introduction.tex
\section{Introduction}
Hexahedral (hex-) meshes are preferred in many mechanical and (bio-)medical engineering applications due to their desired properties for numerical simulations 
\cite{tautges2001generation,gao2017evaluating,BommesLPPSTZ13}.
It is known that the quality of hex-meshes impacts the accuracy and efficiency of various finite element simulations \cite{gao2017evaluating,zhang20053d}. 

\begin{figure}[!t]
\centering
\subfloat[HexaLab]{\includegraphics[width=0.31\columnwidth]{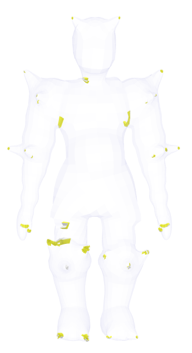}}
\hfill
\subfloat[focus + context volume rendering]{\includegraphics[width=0.31\columnwidth]{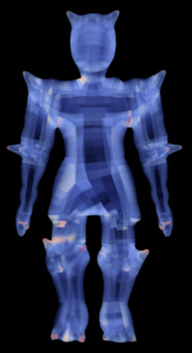}}
\hfill
\subfloat[our method]{\includegraphics[width=0.31\columnwidth]{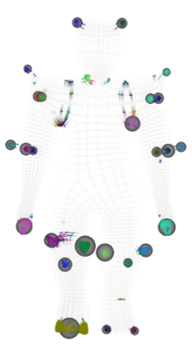}}
\caption{Visualization of the quality of a warrior hex-mesh \cite{livesu2016skeleton} using the quality filtering by the HexaLab \cite{bracci2019hexalab} (a), focus+context volume rendering \cite{neuhauser2021interactive} (b), and our glyph-based method (c), respectively. Our method highlights places with bad-quality elements. The larger the glyphs are, the worse the element quality is. 
}
\label{fig:Motivationcompare}
\end{figure}

Quantitative mesh quality metrics are crucial for mesh generation algorithms. A large number of mesh quality metrics \cite{gao2017evaluating,knupp2006verdict,bracci2019hexalab} have been designed to measure the deviance of individual elements from their ideal shape (e.g., a regular cube for a hexahedral element).
However, existing visualization techniques (see Figure \ref{fig:Motivationcompare}(a)(b)) may overlook small, poor-quality elements due to viewer's attention being drawn to large patterns and areas with large elements. 
Nonetheless, these small and poor-quality elements could lead to the failure of simulations just like the other poor-quality elements.
Boundary preservation is another important criterion impacting the accuracy of boundary condition problems, but most tools lack effective boundary error visualization. Overlapping elements, introduced by some mesh quality improvement algorithms, may also be concealed by traditional color coding or volume rendering methods.

The above limitations of the existing tools motivate our work. To address these limitations, a mesh quality visualization system should provide the following.

\vspace{-0.08in}
\begin{itemize}
  \setlength{\itemsep}{0pt}
  \setlength{\parskip}{0pt}
    \item G1. Effectively reveal regions with bad quality elements despite the region and element sizes. 
    \item G2. Intuitively highlight overlapping/intersecting elements
    \item G3. Show the local configuration of bad elements (e.g., angles at the corners of bad elements) to understand their causes
    \item G4. Display boundary difference/error with respect to the reference surface to highlight places with large differences. 
\end{itemize}
\vspace{-0.08in}
To achieve the above goals, we design and develop a new mesh quality visualization system and make the following contributions.

\vspace{-0.08in}
\begin{itemize}
  \setlength{\itemsep}{0pt}
  \setlength{\parskip}{0pt}
\item We propose a glyph design to highlight places with poor-quality elements independent of the element sizes. This glyph can be aggregated to provide collective information on the quality of elements within a region.
\item We explicitly detect and visualize the overlapping elements that may be hidden in the mesh.
\item We design a comprehensive visualization framework for boundary errors to provide both a summary view of the boundary error statistics and the detailed spatial distribution of error. In particular, parametric representations of 3D boundaries and their errors are used to aid the exploration of the boundary errors for hexahedral meshes.
\item We incorporate the above new design of visualization into an interactive, multi-linked-view visual analytics system to support an effective and efficient study of the mesh quality. 
\end{itemize}
\vspace{-0.08in}

%% file: contents/3-RelatedWord.tex
\section{Related Work}

In this section, we briefly review the literature on hex mesh generation and optimization techniques and the relevant tools that provide the visualization of the meshes.

\noindent \textbf{Hex mesh generation and optimization.}
Hexahedral mesh generation has been a research focus due to its importance in various applications. Earlier techniques ~\cite{LevyL10,lu2001feature,lu2012volumetric,owen2000h,Staten05unconstrainedpaving,staten2010unconstrained,zhang2007adaptive, marechal2009advances,Shepherd2008,sarrate2014unstructured} generated unstructured hex meshes, while recent geometry processing approaches, such as polycube mapping~\cite{HexMeshSGP2011,livesu2013polycut,Li:GPC:13,huang2014,Fang2016AMU,guo2020cut,li2021interactive}, produce semi-structured meshes with limitations in handling complex geometry and topology. 3D parameterization methods~\cite{Huang2011,NieserSGP11,Li2012,Jiang2013}, utilizing orthogonal and non-orthogonal frame fields, have gained popularity but lack guarantees for generating valid hex-meshes. High-quality all-hex mesh generation remains challenging\cite{BommesLPPSTZ13}, and post-processing is often required to improve mesh quality\cite{ji2005global,livesu2015practical,xu2018edgeangle,akram2021embedded, DanielsSSC08,DanielsSC09a, PengZKW11,xu2020semi,gao2015hexahedral,gao2017robust}. 
To support mesh optimization, displaying the obtained meshes' quality is crucial for engineers and researchers to identify areas for improvement.

\noindent \textbf{Meshing tools with visualization capability.}
Due to its importance for various tasks, numerous mesh generation tools have been developed, including generic tools like CUBIT \cite{blacker1994cubit} and other specific tools (or libraries) as listed in \cite{meshgensoftware}. Most of them do not provide a visualization front end to allow the users to check the quality of the generated meshes. A separate tool is usually needed, such as MeshLab \cite{cignoni2008meshlab}, libigl \cite{jacobson2017libigl}, and Paraview \cite{ayachit2015paraview}, to aid the visual analysis of the generated meshes. 
Many finite element simulation softwares \cite{marinkovic2019survey} also offer some mesh generation and processing functionality with limited visualization capability (e.g., wire-frame representation of the meshes, color coding quality values, and color coding different patches). Recently, two works for the visualization and analysis of hexahedral mesh quality have been reported, i.e., the HexaLab \cite{bracci2019hexalab} and the hex mesh structure evaluation and visualization \cite{xu2018hexahedral}. Both works focus on effective visual representation and high-quality rendering to aid the inspection of 3D meshes produced by different methods. However, both tools do not provide split screens to compare two meshes of the same model produced by different techniques. Also, boundary error analysis and overlapping element analysis are not offered by either tool. 
Most recently, a focus+context volume rendering technique \cite{neuhauser2021interactive} has been introduced for the inspection of hexahedral meshes, which allows the user to focus on regions with bad-quality elements without occlusion or cluttering. However, this technique may not reveal bad elements with small sizes (Figure \ref{fig:Motivationcompare} (b)) and does not address the boundary error analysis. In this work, we present a comprehensive visual analysis system that can support the quality and boundary error analysis of hex-meshes.

%% file: contents/4-OurMethod.tex
\section{The Design of HQView}

HQView, a multi-view visual analysis system (Section \ref{sec:multiviews}), offers various representations for 3D hexahedral mesh quality information. It computes a quality metric \cite{knupp2006verdict} for each element corner as a basis for visualization. Besides traditional color encoding and histograms, HQView provides additional functionality to achieve goals G1-G4 listed in the introduction.

\subsection{A Glyph Design for Bad Quality Element (G1)}
\label{sec:glyph}

Element quality is a crucial mesh quality indicator. Traditional color maps work well for 2D meshes but face occlusion issues in 3D hex meshes. Alternative methods like filtering in HexaLab \cite{bracci2019hexalab}, focus+context volume rendering \cite{li2021interactive}, or magic lens techniques have limitations (see \autoref{fig:Motivationcompare}). To achieve size-independent visual encoding, we introduce a glyph design mapping quality values to spheres, with radius encoding the quality value (i.e., the scaled Jacobian metric). 

Instead of using a quality value to describe the regularity of a 3D cell, we calculate the quality of the corners shared by the one-ring neighborhood of a vertex. The smallest quality value is selected to represent the quality of the vertex. The radius of the sphere glyph at each vertex is then determined by the quality value. We define an inverse relationship between the quality of the vertex and the radius of its sphere glyph such that the vertices with bad-quality elements have prominent sphere glyphs that can be easily identified. 
\begin{wrapfigure}{r}{0.25\columnwidth}
\hspace{-0.2in}
\vspace{0.15in}\label{fig:project}\includegraphics[width=0.27\columnwidth]{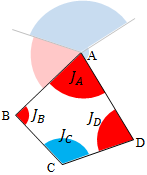}
\end{wrapfigure}
The inset to the right illustrates this mapping where $J_{i}$ represents the scaled Jacobian value at corner $i$.
To support the metric mapping to the radius of the sphere, we map it to the range of $[0, 2]$ as follows.
\begin{equation}
c = 1 - J_{m}
\end{equation}
where $J_{m}$ is the smallest scaled Jacobian of a vertex. $c$ has a reverse relation with $J_{m}$. That is, if $J_{m}=1$ (the best scaled Jacobian), $c=0$, while $c=2$ when $J_{m}=-1$ (the worst scaled Jacobian). The radius of the sphere centered at the vertex is then controlled by $c$. In particular, $r=c*r_{max}$, where $r_{max}$ is a user-controllable parameter to enhance the visibility of the sphere glyphs. Using this strategy, the vertices surrounded by bad-quality elements can be highlighted, regardless of their element sizes. 

\begin{figure}[!t]
\centering
\subfloat[all glyphs]{\includegraphics[width=0.3\columnwidth]{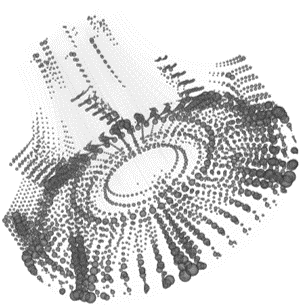}}
\hskip 12pt
\subfloat[aggregated glyphs]{\includegraphics[width=0.3\columnwidth]{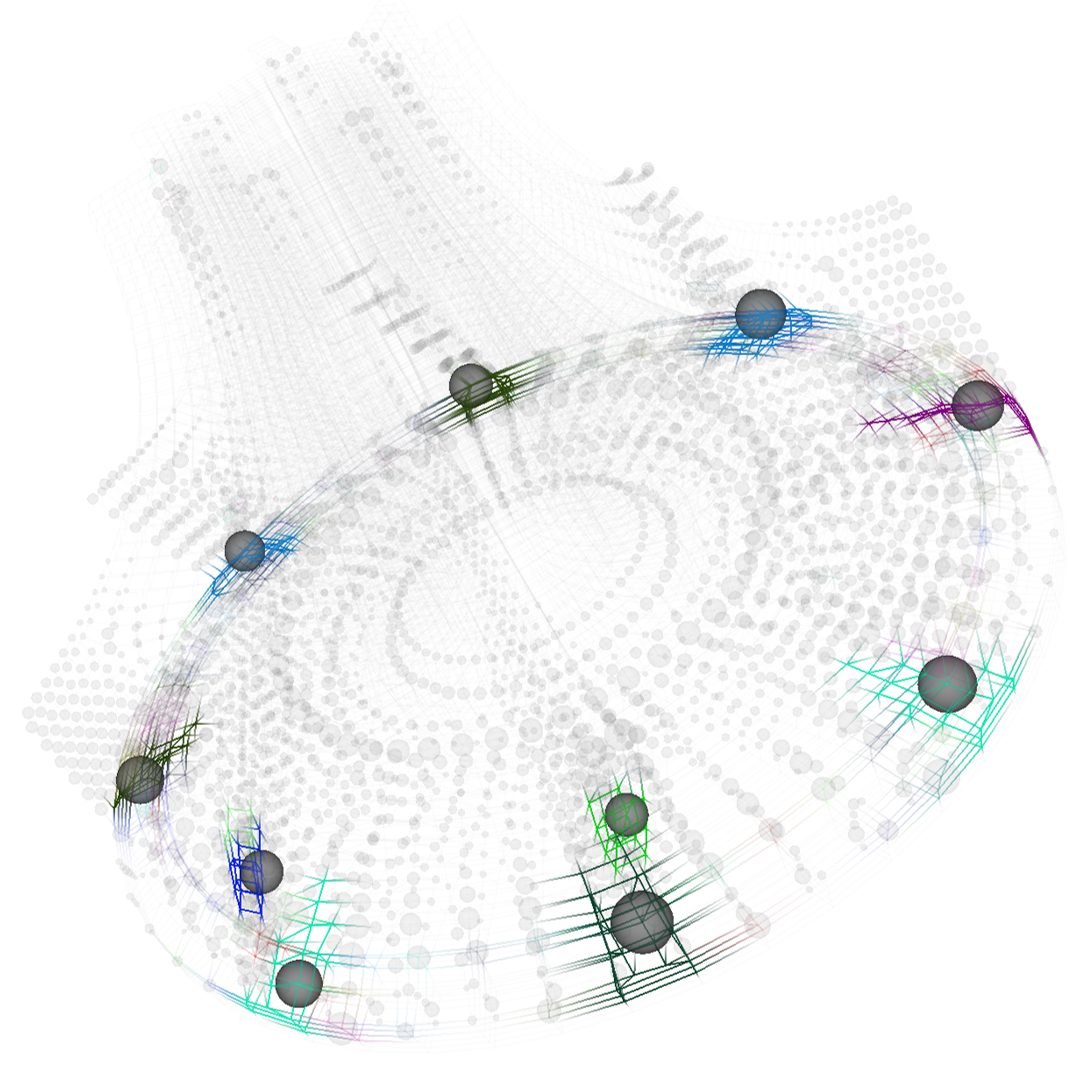}}
\caption{The aggregated glyphs (b) reduce the clutter of non-aggregated glyphs (a).
}
\label{fig:glyphs}
\end{figure}

The above glyph-based visualization can lead to clutter and overlap in areas with many small and bad-quality elements as they lead to large spheres packed in a small region (e.g., \autoref{fig:usecase} (b) and \autoref{fig:glyphs} (a)). In addition, rendering sphere glyphs for all vertices of a large mesh is time-consuming. To address these issues, we introduce a clustering-based glyph aggregating strategy (\autoref{fig:glyphs} (b)).

\begin{figure}[t]
\centering
\includegraphics[width=0.99\linewidth]{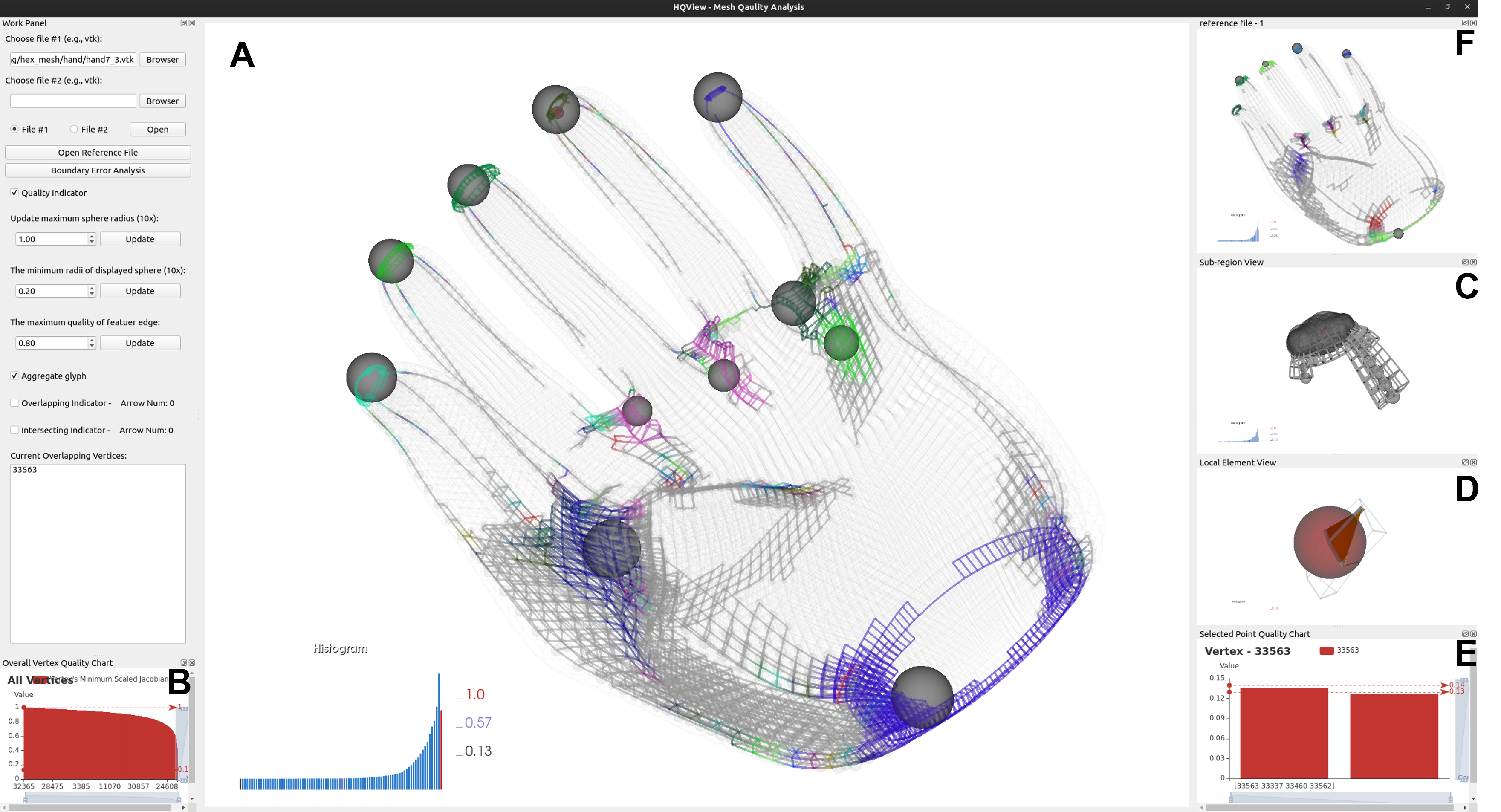}
\caption{The multi-view interface of our system for the level-of-detail inspection of mesh element quality.}
\label{fig:HQViewInterface}
\end{figure}

Our clustering strategy groups nearby vertices based on the spatial proximity of their glyphs. Specifically, if two glyphs overlap (i.e., the sum of their radii exceeds the distance between their centers), the corresponding vertices belong to the same cluster.
The aggregated glyph is positioned at the vertex in the cluster closest to the concentration of cluster vertices. We also employ different colors for edges within distinct clusters. Figure \ref{fig:HQViewInterface} (view A) displays this aggregated visualization, which effectively directs the user's attention to regions with poor-quality elements for further analysis (Section \ref{sec:multiviews}).

\subsection{Overlapping Element Highlighting (G2)}
\label{sec:overlapping}

\begin{figure}[!t]
\centering
\includegraphics[width=.95\columnwidth]{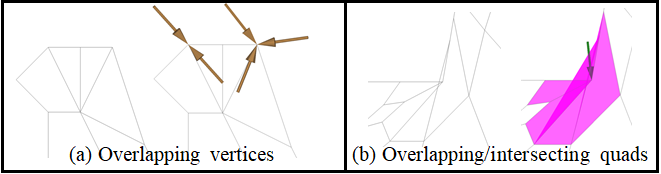}
\caption{Two different overlapping situations may arise (2D illustration): (a) overlapping vertices caused by degenerated edges, (b) overlapping quads caused by one vertex falling inside another quad. Existing visualization (left image of each pair) cannot effectively highlight the overlapping in either case. In contrast, we use arrow(s) to intuitively direct the viewers toward places with overlap (right image of each pair). 
}
\label{fig:overlap_vis}
\end{figure}  


Overlapping elements can be hard to identify, as they may have good quality or be small, making them difficult to distinguish from other bad-quality elements. Two types of overlapping elements can arise (1) overlapping vertices and (2) overlapping cells.

\emph{Overlapping vertices} occur in regions with near-degenerate meshes that don't exhibit inverted elements, like those with zero-length edges. This degenerate configuration is hard to detect and can cause issues during finite element calculations.
\emph{Overlapping cells} are more common and occur when a vertex of a cell moves inside another cell due to mesh operations. These overlapping elements need to be identified and optimized, as they alter calculations in the coordinate system.

Current visualization tools struggle to effectively highlight mesh regions with either overlapping configurations. Our visualization system uses an arrow placement strategy to highlight overlapping vertices (\autoref{fig:overlap_vis}(a)) and transparent shaded rendering for overlapping cells (\autoref{fig:overlap_vis}(b)). 
\revise{In a scenario where a location contains two overlapping vertices, a pair of arrows will be generated, each pointing towards a vertex. If $n$ arrows point to the same vertex, they will be evenly distributed around the vertices at intervals of $360/n$ degrees.}

\subsection{Multi-level Element Quality Analysis (G3)}
\label{sec:multiviews}
Our visual analysis system supports mesh quality inspection at multiple levels. It consists of six views (\autoref{fig:HQViewInterface}): A. main view, B. overall vertex quality chart, C. sub-region view, D. local element view, E. selected point quality chart, and F. reference view.

\noindent \textbf{A. Main View} visualizes mesh quality using glyphs and highlights overlapping elements. It also displays a quality histogram.

\noindent \textbf{B. Overall Vertex Quality Chart} sorts vertices by their quality and supports selection for further analysis.

\noindent \textbf{C. Sub-region View} shows individual elements within clusters for a more detailed study.

\noindent \textbf{D. Local Element View} displays the one-ring neighborhood of a selected vertex, offering a thorough quality analysis.

\noindent \textbf{E. Selected Point Quality Chart} provides quality values of all corners shared by the selected vertex, helping to identify local configuration issues.

\noindent \textbf{F. Reference View} enables comparison of the main mesh with a ground truth or other algorithms' results.

\subsection{Multi-Dimension Views for Boundary Difference Visualization(G4)}

\begin{figure}[t]
\centering
\includegraphics[width=0.8\columnwidth]{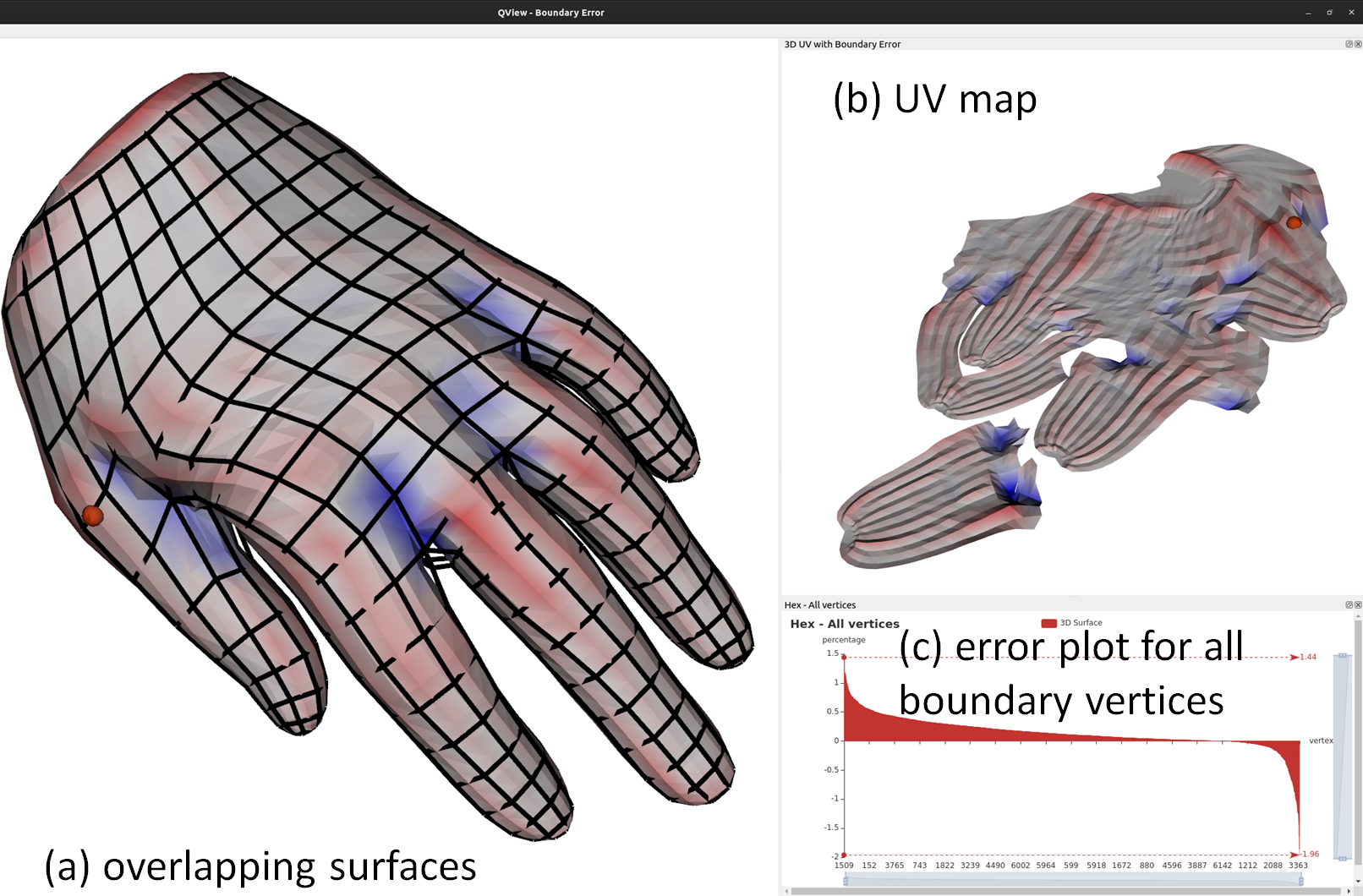}
\caption{A multi-view interface for the boundary error visualization of a hand hex-meshing.}
\label{fig:HexBoundaryError}
\end{figure}

Analyzing the boundary error of 3D meshes is difficult due to occlusion, requiring users to choose different views around the mesh. To efficiently analyze global boundary error information without changing viewpoints, Our system offers a comprehensive boundary error analysis supported by a multi-view visualization for 3D meshes.  

The boundary error is calculated for each surface point based on the closest point on the original mesh. Positive boundary error values (red) indicate modified mesh points lying outside the original mesh, while negative values (blue) indicate points inside the original mesh(Figure \ref{fig:HexBoundaryError} (a)). 
To provide a holistic visualization of the boundary error without requiring the user to rotate the model, we use a UV map of the surface that is extracted from original mesh using the OptCuts \cite{Li:2018:OptCuts}. The UV map unfolds the curve surface to a planar representation.

In the example shown in Figure \ref{fig:HexBoundaryError} (b), positive errors correspond to surface ridges, while negative errors are at concave areas.

A collated percentage graph is provided for all modified mesh surface vertices, offering an overview of the surface areas inside or outside the original mesh. Users can select individual vertices from the graph for further analysis(Figure \ref{fig:HexBoundaryError} (c)).

%% file: contents/6-UserCaseAndEvaluation.tex
\section{Evaluation}
\label{sec:evaluation}

We integrate the above comprehensive visualization techniques into one unified system with two separate windows, i.e., the Mesh Quality Analysis Window, and the Boundary Error Analysis Window. We applied our system to analyze the quality of the hex-meshes included in the database of HexaLab \cite{bracci2019hexalab} and from the hex-mesh structure simplification work \cite{gao2017robust}.

\begin{figure*}[!t]
\centering
\includegraphics[width=0.98\linewidth]{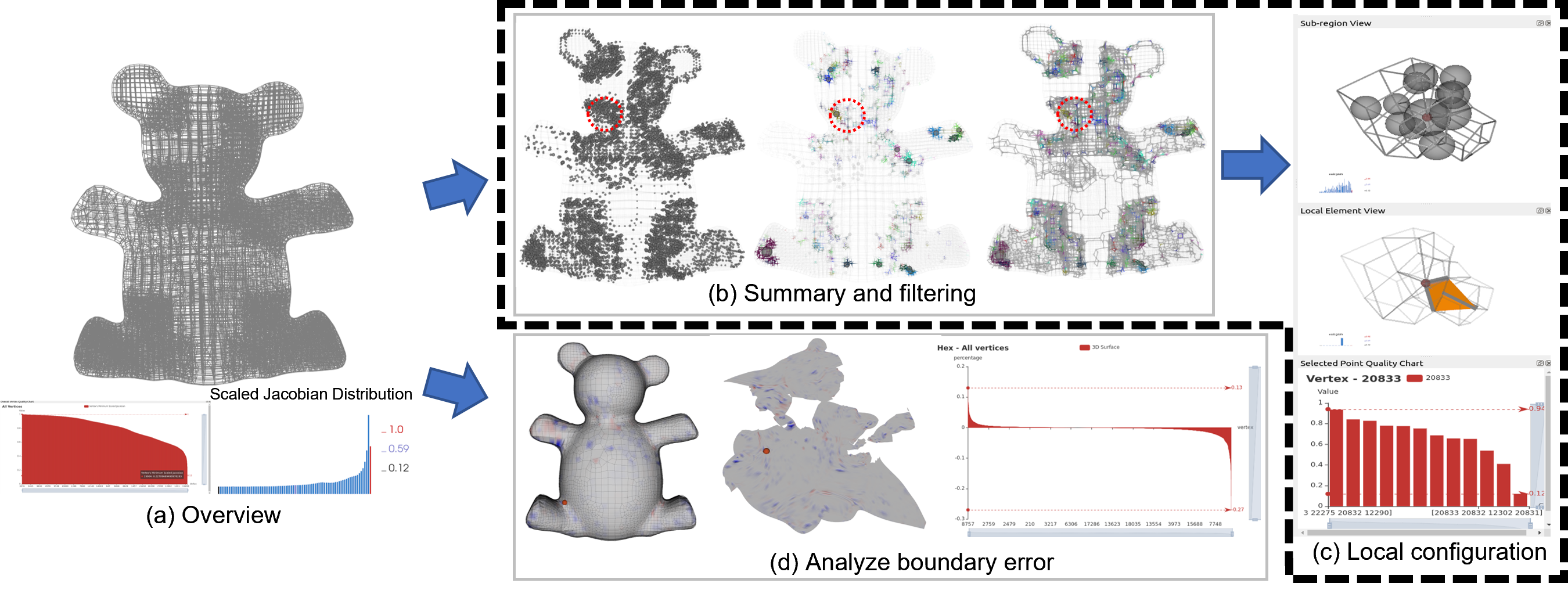}
\caption{A use case of our system for the analysis of a hex-mesh quality.}
\label{fig:usecase}
\end{figure*}

The process of using our system to analyze a hex-mesh is as follows.
After loading a hex-mesh, the Mesh Quality Analysis window displays the wire-frame of the mesh, the distribution of individual element quality, and an interactive bar chart ranking vertex quality (Figure \ref{fig:usecase} (a)). To locate poor-quality elements, glyph representation is activated, using aggregated glyphs to prevent clutter (Figure \ref{fig:usecase}(b)). By focusing on the regions with poor quality, the user can select a large aggregated glyph corresponding to a small cluster and inspect the area in the sub-region view (Figure \ref{fig:usecase}(c), top). A bad-quality vertex can be chosen for detailed inspection, revealing its Jacobian configurations and connectivity configuration (Figure \ref{fig:usecase}(c), middle and bottom). This multi-level, multi-perspective mesh quality analysis process efficiently identifies and analyzes mesh quality issues.

\begin{figure}[th]
\centering
\subfloat[HexaLab]{\includegraphics[width=0.32\columnwidth]{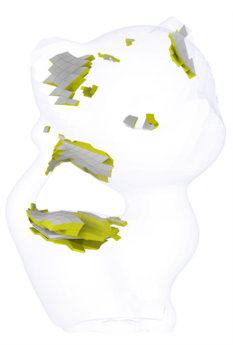}}
\hfill
\subfloat[focus + context volume randering]{\includegraphics[width=0.32\columnwidth]{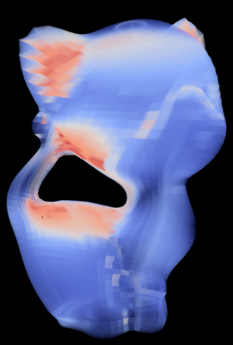}}
\hfill
\subfloat[our method]{\includegraphics[width=0.32\columnwidth]{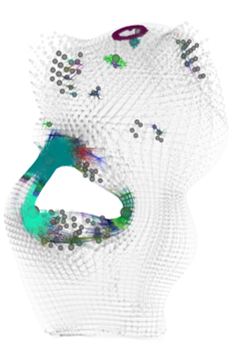}}
\caption{Comparison of the quality visualization using HexaLab (a), focus+context volume rendering (b), and our methods (c) for a teddy bear hex-mesh (top row) and a kitten hex-mesh (bottom), respectively.}
\label{fig:QualityOverviewCompare}
\end{figure}

\vspace{0.05in}
\noindent\textbf{Element quality analysis.}
To evaluate the effectiveness of our proposed visual encoding, we compare our visualizations of a few meshes with those shown by HexaLab \cite{bracci2019hexalab} and the focus+context volume rendering \cite{neuhauser2021interactive}, respectively. Since the other two approaches do not explicitly support the boundary error visualization, our comparison focuses on the element quality analysis and the effective revelation of low-quality elements with different sizes. 
Figure \ref{fig:Motivationcompare} compares the quality visualization of a warrior hex-mesh using the three approaches. Both HexaLab and the focus+context volume rendering cannot effectively reveal the bad elements at the tips of those protruded features. In contrast, our aggregated glyphs not only highlight those places but also convey how severe the element quality is in those regions via the sizes of the glyphs. 

Figure \ref{fig:QualityOverviewCompare} compares the quality visualization of a kitten hex-mesh using the three methods, respectively. From this comparison, we see that among the three approaches, the focus+context volume rendering can provide the smoothest visual representation of the mesh quality. However, if the two areas have a similar quality, the volume rendering will not effectively distinguish their difference as humans cannot accurately tell the difference between similar colors if they are not next to each other (e.g., the back and the left ear of the kitten). Also, one may think the quality of the elements in the left ear of the kitten is worse than those in the right ear because they look more prominent.  In contrast, HexaLab can provide a more accurate reading on the element quality by filtering. That is, the remaining elements after filtering all have quality lower than a user-specified threshold. However, the difference among these remaining elements is hard to discern (e.g., it is hard to decide which elements are worse than the other in Figure \ref{fig:QualityOverviewCompare}(a)). Also, depending on the threshold, other less optimal areas may not be captured (e.g., the top of the kitten and the middle of the tail of the kitten). As a comparison, our aggregated glyph visualization retains most of the areas with bad-quality elements and allows a more effective differentiation of the element quality. For example, the quality of the elements at the back of the neck of the kitten (represented by a big green glyph) is worse than those at the back of the kitty that are prominent in both HexaLab and volume-rendering visualization. Similarly, there is a ring of bad-quality elements at the top of the head of the kitten that is not emphasized by the other two methods due to their small sizes.


\begin{figure}[t]
\centering
\subfloat[Overlapping vertex]{\includegraphics[width=0.45\columnwidth]{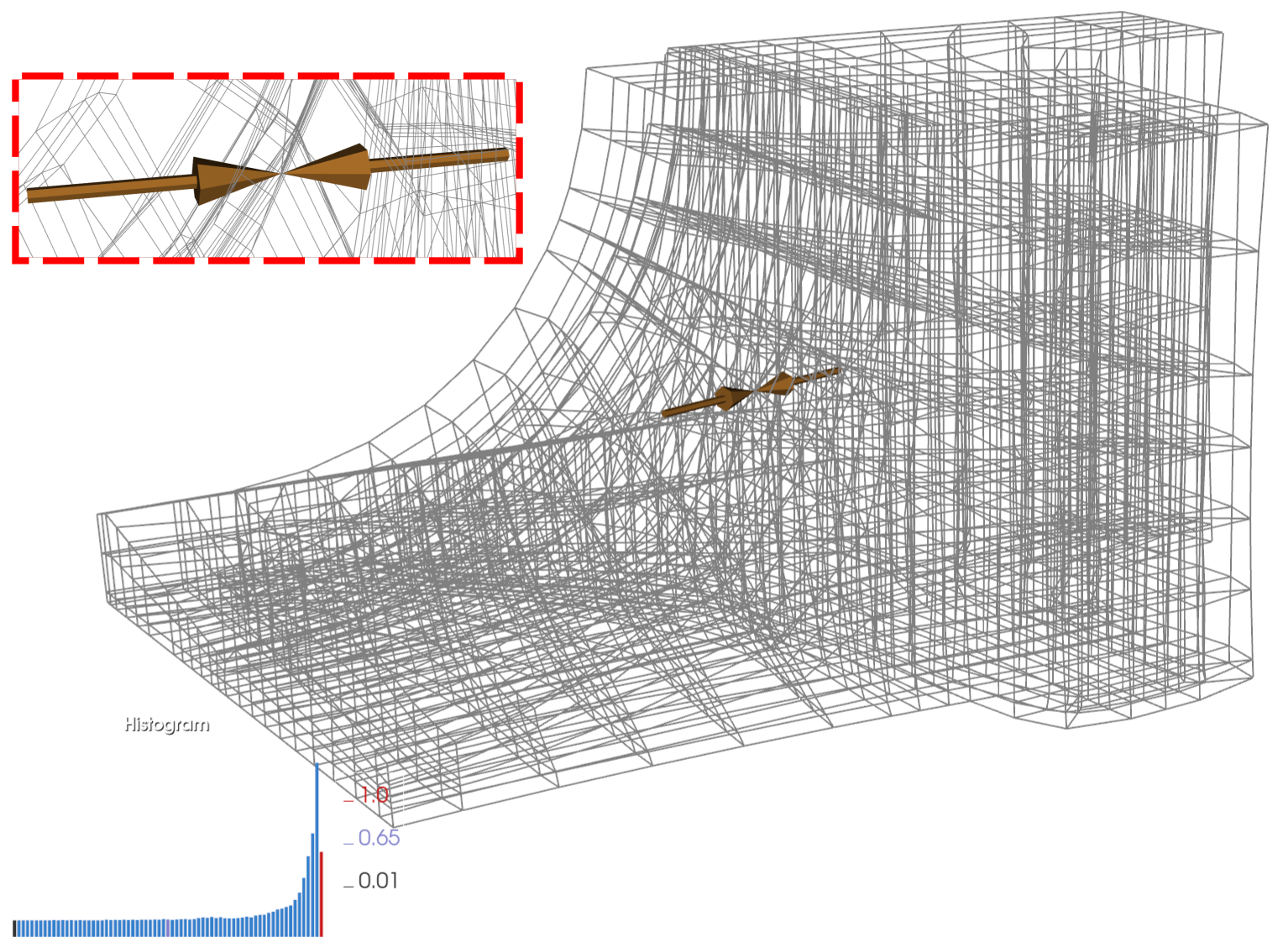}}
\hfill
\subfloat[Intersecting cell]{\includegraphics[width=0.45\columnwidth]{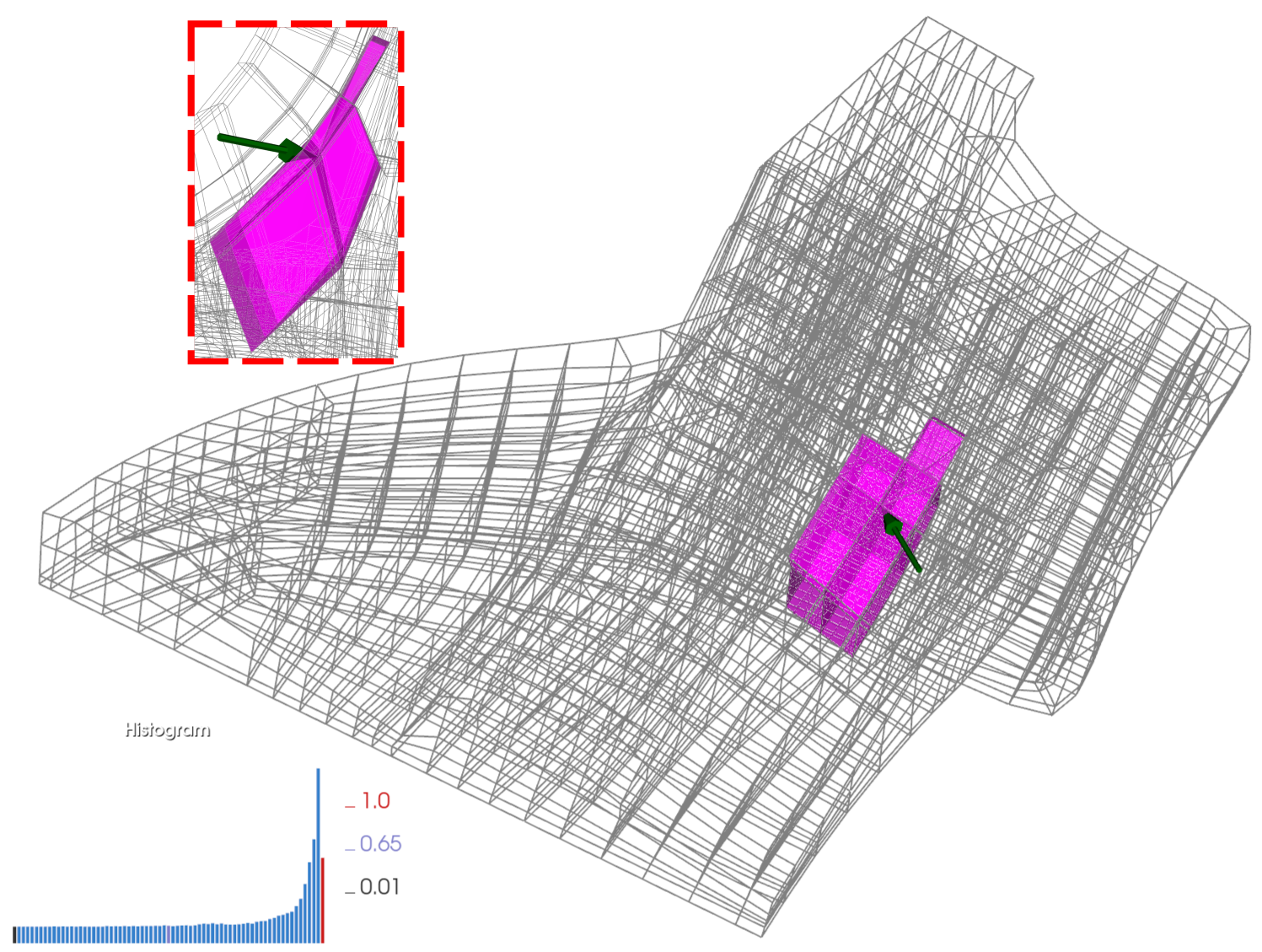}}
\caption{Overlapping vertices and intersecting cells in a 3D view are difficult to be distinguished, as the lines obstruct their visibility.}
\label{fig:intersection_overlapping_3d}
\end{figure}

\vspace{0.05in}
\noindent \textbf{Overlapping cells analysis.} 
\autoref{fig:intersection_overlapping_3d} shows an example of overlapping and intersection elements in a fandisk mesh. The overlapping vertices \autoref{fig:intersection_overlapping_3d}(a) are difficult to notice without the additional arrows. A similar observation can be made for the intersection elements \autoref{fig:intersection_overlapping_3d}(b). In addition, the cells involved in the intersection are highlighted, with each element colored distinctly, and an arrow is placed to indicate each intersection vertex.

\vspace{0.05in}
\noindent\textbf{Boundary error analysis.} To analyze the boundary error of a mesh, the system loads the modified and original meshes into the Boundary Error Analysis Window, as shown in Figure \ref{fig:usecase} (d). In the window, the main view displays overlapping surfaces, but occlusion hides some error distributions. The UV map provides a comprehensive overview without requiring the user to select different viewpoints, though it lacks 3D context like sharp features and corners.
The boundary error distribution doesn't show a consistent pattern near specific feature types. For instance, concave areas corresponding to limb-torso conjunctions can exhibit both positive and negative errors. Larger errors typically occur where the surface has a large curvature, as these areas are challenging to preserve. the sorted error value bar chart enables quick identification of vertices with large boundary errors and their nature. 


\noindent\textbf{User feedback.}
\revise{
We designed an online survey to gather unofficial user feedback on our system. 
Among the 38 responses received, 11 participants identified themselves as mesh experts. 
The survey consists of 13 questions, 
9 of which are designed for mesh element quality analysis, serving three different objectives. The first category of questions requests participants to rank regions by applying their judgment of mesh quality across three different methods. The second category of questions prompts users to identify all problematic areas. The third group of tasks requires users to select the most effective method for highlighting poor-quality regions.
Responses to the questions suggest that our methodology is effective in helping users identify regions of poor quality, particularly when dealing with small mesh sizes.
Detailed feedback of the survey can be found in the supplemental document.
}

%% file: contents/7-Conclusion.tex
\section{Conclusion}
\label{sec:conclusion}

We present a new visual analysis system for the study of the quality of 3D hexahedral meshes. Our system offers simple but effective visual encoding techniques and a multi-view capability to help reveal small elements with low-quality and overlapping configurations and support the inspection of boundary errors.
The evaluation shows that our system outperforms the existing tools in the tasks of locating small elements with low quality, finding overlapping elements in the mesh, and studying boundary error configurations. 

\revise{
To improve our system, we will address the following limitations of the system.
First, our aggregation glyph construction requires performing collision detection among nearby spheres. Our current implementation using traversal has a complexity of $O(n^2)$. To accelerate, we will adopt a pre-computed tree-like structure, such as a union-find data structure \cite{10.1145/62.2160}.
Second, our system does not suggest the ideal configuration for a comparative study of the bad-quality elements. 
Third, our arrow placement strategy for highlighting overlapping elements may still produce cluttered arrows in small regions with many overlapping elements. Nonetheless, the cluttered arrows may help draw the attention of the viewers. To reduce clutter, a view-dependent placement of arrows may be explored. 
Fourth, some models may not be successfully unfolded for boundary error visualization due to the limitation of the used UV unfolding algorithm. 
Finally, the current user evaluation is rather informal and incomplete. Future work will focus on addressing these limitations while considering other element quality measures \cite{gao2017evaluating} and incorporating the visualization of the simulation results run on the respective meshes to provide new insights into the mesh quality and its impact on the downstream tasks.
}